\pdfoutput=1
\documentclass[11pt]{article}
\usepackage[pdftex]{graphicx,color} 
\usepackage{jheppub}
\usepackage{amsmath}
\usepackage{amssymb}
\usepackage{comment}
\usepackage{multirow}
\usepackage{mathtools}
\usepackage{dsdshorthand}
\newcommand{\bea}{\begin{equation}\begin{aligned}}
\newcommand{\eea}[1]{\label{#1}\end{aligned}\end{equation}}
\newcommand{\beq}{\begin{equation}}
\newcommand{\eeq}{\end{equation}}
\newcommand   \xb  {\bar{x}}
\newcommand   \zb  {\bar{z}}
\DeclareMathOperator\Disc{Disc}

\usepackage{tikz}
\usetikzlibrary{arrows,calc,shapes,decorations.pathmorphing,decorations.markings}
\tikzset{
>=stealth',
help lines/.style={dashed, thick},
axis/.style={<->},
important line/.style={thick},
connection/.style={thick, dotted},
  cross/.style={
    cross out,
    draw=black, 
    minimum size=7pt, 
    inner sep=0pt,
    outer sep=0pt
  },
  branchcut/.style={
    decoration={
      snake,
      amplitude=1pt,
      segment length=6pt,
    },
    decorate,
    thick
  },
->-/.style={decoration={
  markings,
  mark=at position #1 with {\arrow{>}}},postaction={decorate}}
}

\title{Dispersion Relation for CFT Four-Point Functions}
\author{Agnese Bissi,}
\author{Parijat Dey,}
\author{Tobias Hansen}
\affiliation{Department of Physics and Astronomy,
	Uppsala University,\\
	Box 516,
	SE-751 20 Uppsala,
	Sweden}
\emailAdd{agnese.bissi@physics.uu.se, parijat.dey@physics.uu.se, tobias.hansen@physics.uu.se}
\preprint{UUITP-42/19}

\abstract{
We present a dispersion relation in conformal field theory which expresses the four point function as an integral over its single discontinuity. Exploiting the analytic properties of the OPE and crossing symmetry of the correlator, we show that in perturbative settings the correlator depends only on the spectrum of the theory, as well as the OPE coefficients of certain low twist operators, and can be reconstructed unambiguously. In contrast to the Lorentzian inversion formula, the validity of the dispersion relation does not assume Regge behavior and is not restricted to the exchange of spinning operators. As an application, the correlator $\langle \phi \phi \phi \phi \rangle$ in $\phi^4$ theory at the Wilson-Fisher fixed point is computed in closed form to order $\epsilon^2$ in the $\epsilon$ expansion.}

\begin{document}
\maketitle

\section{Introduction}
The conformal bootstrap is a tool to study conformal field theories which makes use of the associativity of the operator product expansion (OPE) to constrain the space of possible CFTs.   CFTs can be defined in terms of the operator spectrum of the theory together with the OPE coefficients (CFT data). There have been significant advances in recent years in the analytic domain of the conformal bootstrap, which has produced many results especially in perturbative CFTs.
Such results often involved taking certain limits in the conformal cross-ratios.
For example, the lightcone limit in Lorentzian signature where operators become lightlike separated
is dominated by operators with large spin, and it can be shown that crossing symmetry predicts their CFT data \cite{Fitzpatrick:2012yx,Komargodski:2012ek,Alday:2015eya,Alday:2016njk,Alday:2016jfr}. Crucially, this works to all orders in an expansion in inverse powers of the spin, in practice allowing the full CFT data, up to low spin, to be reconstructed just from singular terms.

More recently in \cite{Caron-Huot:2017vep} it was shown that the CFT data for finite spin operators can be reconstructed from an integral over the double discontinuity of the correlator. This is known as the `inversion formula' and essentially means that it is sufficient to know the singularities of the correlator to compute the CFT data.
It was inspired by scattering theory, where combining the partial wave expansion of the scattering amplitude with the dispersion relation results in the Froissart-Gribov formula \cite{Gribov:1961ex},
which gives the coefficients of the partial wave expansion and makes analyticity in spin manifest.
 Although the inversion formula is enormously successful to compute the CFT data for finite spin operators, there are some limitations which restrict the study of scalar operators in the spectrum. 
Concretely, the derivation of the inversion formula starts with the orthogonality of conformal partial waves and involves a deformation of the integration contour where arcs at infinity only vanish for finite spin, invalidating the derivation for scalar exchanges.
Another way to phrase this is that one encounters a pole when attempting to analytically continue the OPE coefficients along the Regge trajectory down to zero spin.
In addition, since the inversion formula computes the OPE coefficients, the correlator cannot be easily reconstructed in a closed form. In the context of the Wilson-Fisher fixed point in $\phi^4$ theory the inversion formula has been  used  to determine the CFT data for the leading twist operators up to a certain order in perturbation theory \cite{Alday:2017zzv} and the authors propose a method to continue the results down to spin zero, which gives the expected results in this case but is outside the region of validity of the inversion formula. 

One immediate question that arises is  whether it is possible to obtain an integral formula which is not tied to finite spin but is also valid for scalars. In this paper we derive such a relation for CFTs in order to compute the full correlator by exploiting its analytic structure. This complements \cite{Caron-Huot:2017vep} as it is also valid for scalar operators and computes the correlator directly.
The main tool we are using is the dispersion relation adapted to the case of CFTs. 
In the context of scattering theory the dispersion relation gives the scattering amplitude as an integral involving only the discontinuity of the amplitude, which is given by its imaginary part.
We show that the CFT correlator can similarly be obtained as an integral of the single discontinuity of the correlator, exploiting the analytic properties of the OPE as well as crossing symmetry.

Although the dispersion relation is non-perturbative, it will be illustrated in the context of perturbation theory where we have an expansion of the CFT data in terms of the perturbative parameter. We show that the anomalous dimensions of the intermediate operators, together with the OPE coefficients of possible operators with twist below the one of composites of the external operators, determine the full correlator unambiguously.
While the fact that in perturbative settings the singularities are controlled only by the spectrum was already noticeable in the large spin perturbation theory approach \cite{Alday:2016njk},  our formula makes it manifest that the CFT correlator is only sensitive to the operator dimensions and not the OPE coefficients.
In short, perturbative CFT is (mostly) defined by its spectrum.

Our approach is similar in spirit to \cite{Caron-Huot:2017vep}, but there are important differences.
It is clear that the dispersion relation is less constraining than the inversion formula, given that the double discontinuity contains less information than the single discontinuity.
In particular, the inversion formula also constraints anomalous dimensions and not only OPE coefficients.
While the full correlator and the CFT data contain the same information, it is much easier to extract the CFT data from a correlator given in closed form than vice versa. This is an advantage for the dispersion relation which computes directly the correlator.
As to the complexity of the involved integrals, the dispersion relation also tends to be simpler as one integrates the discontinuity only against a single pole as opposed to the conformal block appearing in the inversion formula.
An important point to mention is that this dispersion relation takes into account the intermediate scalar operators without any ambiguity, when the spectrum is fixed. At small coupling this can be thought of as an alternative way of computing the CFT correlators with some inputs using Feynman diagrams. In particular the set of inputs (the spectrum) can be extracted from two point correlation functions which are simpler to compute than three or four point functions.

The paper is organized as follows.
In section \ref{sec:2} we derive the CFT dispersion relation \eqref{eq:bootstrap_dispersion_relation}, based on an analysis of the analytic structure of conformal blocks in $d$ dimensions.
Section \ref{sec:3} shows how the dispersion relation fixes the mean field theory correlator from the presence of the identity operator in the OPE and explains how the anomalous dimensions determine the correlator in a generic perturbative setting.
Section \ref{sec:4} contains two examples of how the dispersion relation can be used in perturbative CFT. The first is the correlator $\< \f \f \f \f \>$ in $\f^4$ theory at the Wilson-Fisher fixed point in $d=4-\e$ dimensions, which is computed in closed form to order $\e^2$ in the $\e$ expansion in \eqref{eq:F2}.
The second example considers $\mathcal{N} = 4$ SYM theory at strong coupling, which holographically corresponds to supergravity in AdS. This example also serves to demonstrate a formula for projecting out  leading order OPE coefficients from the single discontinuity of a correlator.
We conclude in section \ref{sec:5}.

\section{Dispersion relation for CFT }
\label{sec:2}
\subsection{General setup}

Consider the four point function of a scalar operator $\f(x)$ of scaling dimension $\De_\f$
\beq
\< \f (x_1) \f (x_2) \f (x_3) \f (x_4)\> = \frac{F(z,\bar{z})}{(x_1-x_3)^{2\De_\f} (x_2-x_4)^{2\De_\f}}\,,
\eeq
where the right hand side is expressed in terms of the usual cross ratios defined by
\beq
\frac{(x_1-x_2)^{2} (x_3-x_4)^{2}}{(x_1-x_3)^{2} (x_2-x_4)^{2}} =  z \bar{z} =u\,, \qquad
\frac{(x_1-x_4)^{2} (x_2-x_3)^{2}}{(x_1-x_3)^{2} (x_2-x_4)^{2}} =  (1- z) (1- \bar{z})= v\,.
\eeq
The bootstrap equation reads
\beq
F(z,\bar{z}) = F(1-z,1-\bar{z})= \left( z \zb \right)^{-\De_\f} F(1/z,1/\bar{z})\,.
\label{eq:bootstrap}
\eeq
We will refer to the three expressions as $s$-, $t$- and $u$-channel respectively.
The function $F(z, \bar{z})$ can be expanded in the ($s$-channel) conformal blocks
\beq
F(z,\bar{z}) =\left( z \zb \right)^{-\De_\f} \sum\limits_{\De,\ell} a_{\De,\ell} g^{(d)}_{\De,\ell}(z,\bar{z})\,.
\label{eq:OPE}
\eeq
It was shown in \cite{Hogervorst:2013sma} that the sum is convergent on $z, \bar{z} \in \mathbb{C} \setminus (1, +\infty)$.\footnote{What is less obvious is that the OPE acting on the vacuum  also converges in the space of generalized functions when $z$ approaches $(1,+\infty)$ from above or below. This was shown in \cite{Mack:1976pa} for $d=4$. We thank Petr Kravchuk for pointing this out.}

\subsection{Analytic structure of conformal blocks}
\label{sec:blocks}

\begin{figure}
\centering
  \begin{tikzpicture}[scale=1]
    \coordinate (n) at (-1,2.5);
    \coordinate (e) at (3,0);
    \coordinate (w) at (-3,0);
    \coordinate (s) at (-1,-2.5);
    \coordinate (bp1) at (1,0);
    \coordinate (bp2) at (-1,0);
    \draw[->] (w) --  (e) ;
    \draw[->] (s) --  (n) ;
    \filldraw [black] 
     (0,0) circle (2pt) node[below right, black] {$\zb$};
	\draw [branchcut] (w) -- (bp2);
	\draw [branchcut] (2.8,0) -- (bp1);
    \node at (2.8,2.3) [] {$z$};
    \draw[-] (3,2.1) -- (2.6,2.1);
    \draw[-] (2.6,2.1) -- (2.6,2.5);
    \node at (1,0) [cross] {};
    \node at (-1,0) [cross] {};
    \node at (1,-0.5) [] {$1$};
    \node at (-2,0.5) [] {$z^{\frac{\De-\ell}{2}}$};
    \node at (2,0.5) [] {${\tilde g}(z)$};
  \end{tikzpicture}

\caption{Analytic structure of $g^{(d)}_{\De,\ell}(z,\zb)$ and factors contributing to branch cuts.} \label{fig:analytic_structure_4pt}
\end{figure}
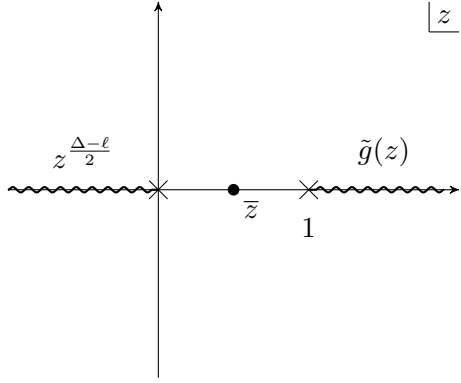

Let us take a closer look at the analytic structure of the conformal blocks. 
The conformal block in $d$ dimensions was computed in \cite{Dolan:2000ut}. It is given by
\beq
g^{(d)}_{\De,\ell}(z,\zb) = (z \zb)^{\frac{\De-\ell}{2}} {\tilde g}^{(d)}_{\De,\ell}(z,\zb)\,,
\label{eq:g_gtilde}
\eeq
where ${\tilde g}$ is given for the exchange of a scalar by\footnote{We allow for four different external scalar operators with $\De_{ij}=\De_i- \De_j$ for a moment because the recursion relation which computes spinning blocks shifts these variables.}
\beq
{\tilde g}^{(d)}_{\De,0}(z,\zb)=\sum\limits_{m,n=0}^\infty
\frac{\left(\frac{\De+\De_{12}}{2} \right)_m \left(\frac{\De-\De_{34}}{2} \right)_m 
\left(\frac{\De-\De_{12}}{2} \right)_{m+n} \left(\frac{\De+\De_{34}}{2} \right)_{m+n} }
{m! n! (\De+1-\frac{d}{2})_m (\De)_{2m+n}} z^m \zb^m (z+\zb -z \zb)^n\,.
\label{eq:gtilde}
\eeq
In order to check for analyticity we have to determine for which values of $z, \zb$ this power series is convergent.
To this end we rewrite \eqref{eq:gtilde} by summing over $n$ which yields a hypergeometric function. This in turn is written in the Euler integral representation in order to finally sum over $m$. The result is
\beq
{\tilde g}^{(d)}_{\De,0}(z,\zb) 
= \frac{\Gamma(\De)}{\Gamma(\frac{\De-\De_{34}}{2}) \Gamma(\frac{\De+\De_{34}}{2})} \int\limits_0^1 dt
\frac{{}_2 F_1 \left(\tfrac{\De-\De_{12}}{2},\tfrac{\De+\De_{12}}{2},1-\tfrac{d}{2}+\De, \tfrac{t(1-t)z \zb}{1-t(z+\zb-z \zb)} \right)}{
t^{1-\frac{\De+\De_{34}}{2}}
(1-t)^{1-\frac{\De-\De_{34}}{2}}
(1-t(z+\zb-z \zb))^{\frac{\De-\De_{12}}{2}}
}\,.
\label{eq:gtilde_integral}
\eeq
All we have to do now is to check for finiteness of this integral. For $\De > \max (|\De_{12}|,|\De_{34}|)$ the integrand has the following singularities
\bea
1 &= \frac{t(1-t)z \zb}{1-t(z+\zb-z \zb)} \qquad &\Leftrightarrow \qquad &t \in \left\{ \frac{1}{z}, \frac{1}{\zb} \right\}\,,\\
0 &= 1-t(z+\zb-z \zb) \qquad &\Leftrightarrow \qquad &t = \frac{1}{1-(1-z)(1-\zb)}\,.
\eea{eq:gtilde_singularities}
We conclude that the integral is finite and hence ${\tilde g}^{(d)}_{\De,0}(z,\zb)$ is analytic when all the singularities lie outside the integration region i.e.\ for
\beq
z, \bar{z} \in \mathbb{C} \setminus (1, +\infty) \quad \text{with} \quad (1-z)(1-\zb) \in \mathbb{C} \setminus  (-\infty,0)\,.
\eeq
Since it will be enough to exploit analyticity in one variable, we will set $\zb$ to some fixed value between 0 and 1. In this case the domain of analyticity in $z$ becomes
\beq
z \in \mathbb{C} \setminus (1, +\infty)\,.
\eeq
The ${\tilde g}$ for the exchange of spinning operators are given in terms of \eqref{eq:gtilde} by a recursion relation \cite{Dolan:2000ut} in $\ell$, hence they inherit the same analytic properties.
We conclude that the conformal blocks have the analytic structure depicted in Figure \ref{fig:analytic_structure_4pt}. There is a branch cut for $z<0$ that originates from the generically non-integer power of $z$ and another branch cut for $z>1$ from ${\tilde g}$.
Naturally, the branch cut from the overall power is much simpler and it is easy to compute the discontinuity
\beq
\underset{z\, <\, 0}{\Disc}\, (z \zb)^{-\De_\f}   g^{(d)}_{\De,\ell}(z,\bar{z}) = \left(1-e^{2\pi i \left(\De_\phi - \frac{\De- \ell}{2}\right) }\right) (z \zb)^{-\De_\f} g^{(d)}_{\De,\ell}(z,\bar{z})\,,
\label{eq:disc_sin}
\eeq
where the discontinuity is defined as
\beq
\underset{z}{\Disc}\, f(z) \equiv \lim_{\a \to 0^+} f(z + i \a) - f(z - i \a) \,.
\eeq
For later reference we also note down the conformal blocks for the case $d=4$
\bea
g^{(4)}_{\De,\ell}(z,\bar{z}) ={}& \frac{z \bar{z}}{z-\bar{z}} \left( k_{\frac{\De+\ell}{2}} (z) k_{\frac{\De-\ell-2}{2}} (\bar{z})-k_{\frac{\De-\ell-2}{2}} (z) k_{\frac{\De+\ell}{2}} (\bar{z}) \right)\,,
\eea{eq:4d_block}
where
\beq
k_\b (z) = z^{\b} {}_2 F_1 \left( \b,\b,2\b,z \right)\,.
\label{eq:k}
\eeq

\subsection{Dispersion relation}
\label{sec:dispersion_relation}

\begin{figure}
\centering
  \begin{tikzpicture}[scale=1.8]
    \coordinate (n) at (-0,2);
    \coordinate (e) at (2,0);
    \coordinate (w) at (-2,0);
    \coordinate (s) at (-0,-2);
    \coordinate (bp1) at (1,0);
    \coordinate (bp2) at (0,0);
    \draw[->] (w) --  (e) ;
    \draw[->] (s) --  (n) ;
    \filldraw [black] (0.33,0) circle (1pt);
    \filldraw [black] (0.67,0) circle (1pt);
	\draw [branchcut] (w) -- (bp2);
	\draw [branchcut] (1.9,0) -- (bp1);
    \node at (1,0) [cross] {};
    \node at (-0,0) [cross] {};
    \node at (0.33,-0.25) [] {$z$};
    \node at (0.67,-0.25) [] {$\zb$};
    \node at (1,-0.25) [] {$1$};
    \node at (1.85,1.85) [] {$z'$};
    \draw[-] (2,1.7) -- (1.7,1.7);
    \draw[-] (1.7,1.7) -- (1.7,2);
    \draw[->-=.33] (2,0.1) arc (0:180:2);
    \draw[->-=.5] (-2,0.1) -- (0,0.1);
    \draw[-] (0,0.1) arc (90:-90:0.1);
    \draw[->-=.5] (0,-0.1) -- (-2,-0.1);
    \draw[->-=.33] (-2,-0.1) arc (180:360:2);
    \draw[->-=.5] (2,-0.1) -- (1,-0.1);
    \draw[-] (1,-0.1) arc (270:90:0.1);
    \draw[->-=.5] (1,0.1) -- (2,0.1);
    \draw[->-=.5] (0.33,-0.1) arc (-90:270:0.1);
  \end{tikzpicture}
\caption{Contour deformation determining the dispersion relation.} \label{fig:contour_deformation}
\end{figure}
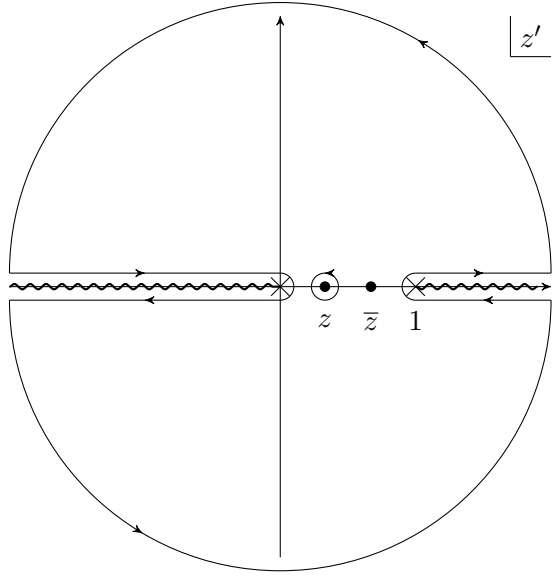

The term dispersion relation in the context of QFT is usually used for a relation expressing scattering amplitudes in terms of their imaginary part. Here we mean the same mathematical relation applied to CFT correlation functions. The idea is simply to introduce a pole at a generic point $z'$ and write the correlator as the residue of this pole
\beq
F(z,\zb) = \frac{1}{2 \pi i} \oint\limits_{z} dz' \, \frac{1}{z' - z} F(z',\zb)\,.
\label{eq:dispersion_relation_ansatz}
\eeq
Next the contour can be deformed as illustrated in figure \ref{fig:contour_deformation}. The analytic properties of the correlator follow from the analytic properties of the conformal blocks through the OPE \eqref{eq:OPE}.
As a result the contour can be deformed to wrap around the branch cuts on the real axis. 
To determine the contribution from the arcs at infinity we have to look at the Laurent series of $F(z,\zb)$ at $z=\infty$ which is given by the $u$-channel OPE
\beq
F(z,\bar{z}) =\sum\limits_{\De,\ell} a_{\De,\ell} g^{(d)}_{\De,\ell}(1/z,1/\bar{z})\,.
\eeq
Due to the overall factor of $z^{\frac{\De-\ell}{2}}$ in the conformal block \eqref{eq:g_gtilde}, only primaries with $\De-\ell \leq 0$ can contribute to the arc at infinity. For unitary theories and $d>2$ the only such operator is the identity and its arc contribution is simply 1.
We end up with the following dispersion relation, expressing the correlator in terms of its discontinuity
\beq
F(z,\zb) = 1 + \frac{1}{2 \pi i} \int\limits_{-\infty}^\infty dz' \, \frac{1}{z' - z} \underset{z'}{\Disc}\, F(z',\zb)\,.
\label{eq:dispersion_relation}
\eeq
We just saw in section \ref{sec:blocks} that the discontinuity of the conformal block at $z<0$ is dramatically simpler than the one at $z>1$. Luckily, when it comes to correlators we can use crossing symmetry \eqref{eq:bootstrap} to express the complicated discontinuity in terms of the simple one
\beq
\underset{z\, >\, 1}{\Disc}\, F(z,\zb) 
= - \underset{z\, <\, 0}{\Disc}\, F(1-z,1-\zb) \Big|_{\substack{z\to 1-z\\\zb \to 1-\zb}} 
=- \underset{z\, <\, 0}{\Disc}\, F(z,\zb) \Big|_{\substack{z\to 1-z\\\zb \to 1-\zb}} \,.
\eeq
This can be used to rewrite the integral over the branch cut on the positive real axis
\beq
\int\limits_{1}^\infty dz' \, \frac{1}{z' - z} \underset{z'}{\Disc}\, F(z',\zb)
= \int\limits_{-\infty}^0 dz' \, \frac{1}{z' - (1-z)} \underset{z'}{\Disc}\, F(z',1-\zb)\,,
\eeq
immediately leading to a simplified version of the dispersion relation where we integrate only over one branch cut
\beq
F(z,\zb) = 1 + \left(\frac{1}{2 \pi i} \int\limits_{-\infty}^0 dz' \, \frac{1}{z' - z} \underset{z'}{\Disc}\, F(z',\zb) + (z,\zb) \to (1-z,1-\zb)\right) \,.
\label{eq:bootstrap_dispersion_relation}
\eeq

\section{Computing correlators}
\label{sec:3}

Let us now put the dispersion relation to use and compute some correlators. We begin by showing how the presence of the identity operator in the OPE leads to the familiar mean field theory correlator. Then we discuss how to compute correlators in situations that allow for an expansion in a small parameter.

To see how the presence of the identity in the OPE generates the correlator, consider the $s$-channel identity
\beq
\frac{1}{(z \zb)^{\De_\f}}\,.
\eeq
Its discontinuity is
\beq
\underset{z\, <\, 0}{\Disc}\, (z \zb)^{-\De_\f}   = (1-e^{2\pi i \De_\phi}) (z \zb)^{-\De_\f}\,.
\label{eq:disc_id}
\eeq
In order to use the dispersion relation we need the integral
\beq
\int\limits_{-\infty}^0 dz' \, \frac{(z')^{-\De_\f} }{z' - z} = 2 \pi i \frac{z^{-\De_\f}}{1-e^{2\pi i \De_\phi}}\,, \qquad 0< \De_\f < 1\,,
\eeq
and hence the dispersion relation \eqref{eq:bootstrap_dispersion_relation} yields the familiar mean field theory correlator
\beq
F^{MF}(z,\zb) = 1 + \frac{1}{(z \zb)^{\De_\f}} + \frac{1}{((1-z)(1- \zb))^{\De_\f}}\,.
\label{eq:F0}
\eeq
It is interesting to note that the argument can be also repeated separately for the case $\De_\f \in \mathbb{N}$ when the prefactor in \eqref{eq:disc_id} vanishes. In this case the discontinuity is a delta function
\bea
\underset{z}{\Disc}\, (z \zb)^{-\De_\f}
={}&  \zb^{-\De_\f} \frac{2 \pi i (-1)^{\De_\f}}{(\De_\f-1)!} \partial_z^{\De_\f-1} \de(z) \,,
\eea{eq:DF_from_id}
and the dispersion relation yields the same result.
Here we used that the discontinuity of a simple pole is given by
\beq
\Disc \frac{1}{z} = -2 \pi i \de(z)\,,
\label{eq:disc_pole}
\eeq
as can be seen by integrating against a test function
\beq
 \int_{-\infty}^\infty dz \, f(z) \Disc \frac{1}{z} =
\lim_{\a \to 0^+} \int_{-\infty}^\infty dz \, f(z) \left(\frac{1}{z+i \a} - \frac{1}{z-i \a} \right) = -2 \pi i f(0)\,,
\eeq
where we commuted limit and integral to arrive at the result.
The discontinuities of higher poles are obtained by acting with derivatives on \eqref{eq:disc_pole}.
To conclude this section, let us cite the decomposition of the correlator \eqref{eq:F0} into conformal blocks
\beq
F^{MF}(z,\zb) = \frac{1}{\left( z \zb \right)^{\De_\f}} \left( 1+ \sum\limits_{n=0}^{\infty} \sum\limits_{\substack{\ell=0\\\text{even}}}^{\infty} a^{MF}_{n,\ell} \, g^{(d)}_{2 \De_\f +2n+\ell,\ell}(z,\zb)\right)\,,
\eeq
with the coefficients \cite{Fitzpatrick:2011dm}
\beq
a^{MF}_{n,\ell} = \frac{(1+(-1)^\ell) (\De_\f-\frac{d}{2}+1)_n^2 (\De_\f)_{n+\ell}^2}{\ell!n! (\ell+\frac{d}{2})_n (2\De_\f+n-d+1)_n (2\De_\f+2n+\ell-1)_\ell (2\De_\f+n+\ell-\frac{d}{2})_n}\,.
\label{eq:a0}
\eeq

\subsection{Perturbative correlators}

Next let us discuss correlators that depend on a small parameter $\e$.
For concreteness we discuss expansions around the mean field theory correlator discussed in the previous section.
That means that both the anomalous dimensions and the OPE coefficients will receive corrections proportional to $\e$
\bea
\De_{n,\ell} &= 2 \De_\f +2n+ \ell + \e \gamma^{(1)}_{n,\ell}  + \e^2 \gamma^{(2)}_{n,\ell}+O(\e^3)\,,\\
a_{n,\ell} &= a_{n,\ell}^{MF} + \e a_{n,\ell}^{(1)} + \e^2 a_{n,\ell}^{(2)}+O(\e^3)\,,
\eea{eq:FT_data}
which leads to a correction of the correlator
\beq
F(z,\zb) = F^{MF} (z,\zb) + \e F^{(1)} (z,\zb)  + \e^2 F^{(2)} (z,\zb)+O(\e^3)\,.
\eeq
The leading contribution is given by
\bea
F^{(1)} (z,\zb) &= (z \zb)^{-\De_\f} \sum\limits_{n=0}^\infty \sum\limits_{\substack{\ell=0\\\text{even}}}^\infty 
\left( a_{n,\ell}^{(1)} + a_{n,\ell}^{(0)} \gamma^{(1)}_{n,\ell} \partial_{\e} \right) \, g^{(d)}_{2\De_\f+2n+\ell,\ell}(z,\bar{z})\,,
\eea{eq:HG_def_4D}
however one sees from \eqref{eq:disc_sin} that the discontinuity at $z<0$ does not depend on $a_{n,\ell}^{(1)}$ 
\beq
\underset{z\, <\, 0}{\Disc}\, F^{(1)} (z,\zb) = \pi i (z \zb)^{-\De_\f} \sum\limits_{n=0}^\infty \sum\limits_{\substack{\ell=0\\\text{even}}}^\infty 
a_{n,\ell}^{(0)} \gamma^{(1)}_{n,\ell}  \, g^{(d)}_{2\De_\f+2n+\ell,\ell}(z,\bar{z})\,.
\label{eq:DF_from_gamma}
\eeq
By comparing to \eqref{eq:HG_def_4D} one sees that the only contribution to this discontinuity is generated when the derivative in $\e$ acts on powers of $z$, i.e.\ $\partial_\e z^{\frac{\e}{2}} = \frac{1}{2} \log (z) z^{\frac{\e}{2}}$.
Note that if there are operators with bare dimension $\De-\ell = 2 \De_\f - n, n =1,2,\ldots$, there are further delta function singularities like the one from the identity. This is not the case in the current setup, but it will be in the example of section \ref{sec:sugra}.
Since the discontinuity \eqref{eq:DF_from_gamma} determines the correlator via the dispersion relation \eqref{eq:bootstrap_dispersion_relation}, it follows that the four-point function and hence the OPE coefficients $a^{(1)}_{n,\ell}$ are entirely defined by the spectrum of the CFT and OPE coefficients of possible operators with twist below $2 \De_\f$! This statement holds generically in situations where a small parameter is present and makes manifest the fact that the spectrum together with crossing symmetry is enough to specify the CFT. 
For higher orders in the perturbation the OPE coefficients of the previous order become an input at the next order and determine the new OPE coefficients.

\section{Applications}
\label{sec:4}
\subsection{Wilson-Fisher model}

To study a concrete example, we will now consider the correlator $\langle \f \f \f \f \rangle$
in $\f^4$ theory at the Wilson-Fisher fixed point in an $\epsilon$-expansion.
This model has recently been studied using the Lorentzian inversion formula \cite{Alday:2017zzv} and serves well to illustrate the difference between the two methods. To start with, being based on the double discontinuity the Lorentzian inversion formula can generally reach one order higher in the expansion with similar input. With the discontinuity one has to analyze all terms containing $\log (z)$ whereas the double discontinuity extracts $\log^2 (z)$ and higher powers of logarithms. On the other hand, the dispersion relation yields closed expressions for the correlator, compared with the inversion formula which computes OPE coefficients. Summing the OPE to obtain the correlator from CFT data is generically hard and we are not aware that it has been done for the example at hand. Furthermore, in \cite{Alday:2017zzv} the correct OPE coefficients of scalars have been obtained by performing an analytic continuation around a pole in $\ell$, but it is not clear to what extent this was justified given that the Lorentzian inversion formula cannot be safely used for scalars.
We will perform the $\e$ expansion where both the CFT data and the dimension
\beq
d=4-\e\,,
\eeq
depend on the expansion parameter. The starting point is the mean field theory correlator \eqref{eq:F0}
where we insert the appropriate external dimension
\beq
\De_\f = 1 - \frac{1}{2} \e +\frac{1}{108} \e^2+ O(\e^3)\,.
\eeq
The OPE coefficients \eqref{eq:a0} with $n \neq 0$ vanish up to $O(\e)$
\beq
a^{MF}_{n,\ell} = \frac{(1+(-1)^\ell) (\De_\f)_\ell^2}{\ell! (2 \De_\f +\ell-1)_\ell} \de_{n,0}+ O(\e^2) \,,
\eeq
which means that at leading order only the identity and the double trace operators of twist approximately two appear in the OPE
\beq
\f \times \f = 1 + \f \partial^\ell \phi + O(\e)\,.
\eeq
In the following we will omit the index $n$ of $a_{n,\ell}$ and $\De_{n,\ell}$ whenever it is zero.
The only missing ingredient to compute the discontinuity \eqref{eq:DF_from_gamma} are the dimensions of these operators (see for instance \cite{Gopakumar:2016cpb})
\bea
\De_{0} &= 2 \De_\f + \frac{1}{3} \e + \frac{8}{81} \e^2+O(\e^3)\,,\\
\De_{\ell} &= 2 \De_\f + \ell - \frac{1}{9 \ell (\ell+1)} \e^2+O(\e^3)\,, \qquad \ell >0.
\eea{eq:dimensions_twist2}
Since the spinning operators have no anomalous dimension at order $\e$, the only operator contributing to the discontinuity is $\f^2$
\beq
\underset{z\, <\, 0}{\Disc}\, F^{(1)} (z,\zb) = \frac{2}{3} \pi i (z \zb)^{-1} g_{2,0}(z,\bar{z}) = 
\frac{2}{3}  \pi i \frac{\log(1-\zb) - \log(1-z)}{z-\zb}\,.
\eeq
Moreover, there are no operators in the $s$-channel OPE apart from the identity that contribute to poles at $z=0$. Since the contribution of the identity is fully accounted for by the mean field theory correlator, there will be no further delta function singularities. 
Using the dispersion relation \eqref{eq:bootstrap_dispersion_relation} (without the 1 which only contributes at $O(\e^0)$)
we find
\beq
F^{(1)} (z,\zb) = \frac{1}{3(z-\bar{z})}\left(\log(z \bar{z}) \log \left( \frac{1-\zb}{1-z}\right) -2 \text{Li}_2 (z)+2 \text{Li}_2 (\bar{z}) \right)\,.
\label{eq:F1}
\eeq
Now we can expand this result into conformal blocks and find that only the OPE coefficient of $\f^2$ is corrected
\beq
a_{n,\ell}^{(1)} = -\frac{2}{3} \de_{n,0} \de_{\ell,0}\,.
\eeq
To compute the correlator at the next order we simply expand the correlator in $\e$ and keep only the terms containing a $\log(z \zb)$
\bea
F(z, \zb) ={}& a_{0} \left( \frac{1}{2}(\gamma^{(1)}_{0} \e + \gamma^{(2)}_{0} \e^2) \log(z \zb) (1+\e \partial_\e) + \frac{1}{8} (\gamma^{(1)}_{0})^2 \e^2 \log(z \zb)^2 \right) {\tilde g}^{(4-\e)}_{\De_{0},0} (z,\zb)\\
&+ \sum\limits_{\substack{\ell=2\\\text{even}}}^\infty 
\frac{1}{2} a_{\ell} \gamma^{(2)}_{\ell} \e^2 \log(z \zb) {\tilde g}^{(4-\e)}_{\De_{\ell},\ell} (z,\zb)
+ \text{continuous at } z<0\,.
\eea{eq:F_eps2_logs}
The discontinuity at second order in $\e$ follows straightforwardly to be
\begin{align}
\underset{z\, <\, 0}{\Disc}\, F^{(2)} (z,\zb) ={}& \pi i 
\left( a^{(1)}_{0} \gamma^{(1)}_{0} +  a^{MF}_{0} \gamma^{(2)}_{0} + \tfrac{1}{2}  a^{MF}_{0} (\gamma^{(1)}_{0})^2 \log(-z \zb) +  a^{MF}_{0} \gamma^{(1)}_{0} \partial_\e \right) {\tilde g}^{(4)}_{2,0} (z,\zb) \nonumber\\
& + 2 \pi i \sum\limits_{\substack{\ell=2\\\text{even}}}^\infty \frac{\Gamma(\ell+1)^2}{\Gamma(2\ell+1)}
\gamma^{(2)}_{\ell} {\tilde g}^{(4)}_{2+\ell,\ell} (z,\zb)\,.
\label{eq:disc_F2}
\end{align}
The sum in the second line can be done using that the conformal blocks for twist two operators in four dimensions \eqref{eq:4d_block} simplify
\beq
{\tilde g}^{(4)}_{2+\ell,\ell} (z,\zb) = \frac{k_{\ell+1}(z) - k_{\ell+1}(\zb)}{z-\zb}\,,
\eeq
and\footnote{Use the integral representation ${}_2 F_1 \left(a,b,c,z \right) = \frac{\Gamma(c)}{\Gamma(b)\Gamma(c-b)}
\int\limits_0^1 dt\, t^{b-1} (1-t)^{c-b-1} (1-t z)^{-a}$ in order to sum.}
\beq
\sum\limits_{\substack{\ell=2\\\text{even}}}^\infty \frac{\Gamma(\ell+1)^2}{\Gamma(2\ell+1) \ell (\ell+1)}
k_{\ell+1}(z)
=\log(1-z) +\frac{1}{4} \log(1-z)^2 + \text{Li}_2(z)\,.
\label{eq:spin_blocks_sum}
\eeq
The most complicated part of the discontinuity comes from the expansion of the conformal block, which has to be done using the expression for general dimensions in \eqref{eq:gtilde}\footnote{Expand the integrand of \eqref{eq:gtilde_integral} in $\e$ (for instance using HypExp \cite{Huber:2005yg,Huber:2007dx}) before doing the integral.}
\begin{align}
{}&\partial_\e {\tilde g}^{(4)}_{2,0} (z,\zb) =
\frac{1}{z-\zb} \left(\frac{2}{3} \left(\text{Li}_2(z) -\text{Li}_2(\zb)\right) 
+\frac{1}{2} \left( \text{Li}_2 \hspace{-2pt} \left( \tfrac{\zb}{z} \right) -\text{Li}_2\hspace{-2pt} \left( \tfrac{z}{\zb} \right) + \text{Li}_2\hspace{-2pt} \left( \tfrac{z(1-\zb)}{\zb(1-z)} \right) -\text{Li}_2\hspace{-2pt} \left( \tfrac{\zb(1-z)}{z(1-\zb)} \right)
 \right)
\right.\nonumber\\
&+\left.
\frac{1}{2}\log\left(\tfrac{1-z}{1-\zb} \right)
\left(
\frac{4}{3} - \log(z-\zb)- \log(\zb-z)+ \log (z \zb) +\frac{1}{2} \log ((1-z)(1-\zb))
\right)
\right)\,.
\label{eq:gtilde_eps}
\end{align}
Inserting the discontinuity into the dispersion relation one can do the integral and finds the correlator\footnote{The most convenient way to do the integral seems to be to use the Maple package HyperInt \cite{Panzer:2014caa}.}
\begin{align}
&F^{(2)} (z,\zb) = \frac{1}{z-\zb} \bigg[
-\frac{1}{12} \text{log} \left(\tfrac{z}{\zb}\right) \text{log} ^2\left(\tfrac{1-z}{1-\zb}\right)
-\frac{1}{12} \text{log} ^2((1-z) (1-\zb)) \text{log} \left(\tfrac{z}{\zb}\right)
\nonumber\\
&+\text{log} \left( \tfrac{1-z}{1-\zb} \right) \left(
\frac{10}{81} \text{log} \left( z \zb \right) 
+\frac{1}{12} \text{log}^2\left(\tfrac{z}{\zb}\right) 
-\frac{1}{36} \text{log} ^2(z \zb) 
-\frac{1}{9} \text{log} ((1-z) (1-\zb)) \text{log} (z \zb)
\right)
\nonumber\\
&-\frac{1}{18}
   (\text{Li}_2(z)-\text{Li}_2(\zb)) \left(4 \text{ log} ((1-z) z)+4 \text{ log} ((1-\zb) \zb)-\frac{40}{9}\right)
\label{eq:F2}\\
&+\frac{1}{3} \left( \left(\text{Li}_2\left(\tfrac{\zb-z}{\zb-1}\right) +\frac{1}{4} \text{log} ^2\left(\tfrac{1-z}{1-\zb}\right)\right) \text{log} (z
   \zb)
-\left(\text{Li}_2\left(\tfrac{\zb-z}{\zb}\right) +\frac{1}{4} \text{log}^2\left(\frac{z}{\zb}\right)\right)\text{log} ((1-z) (1-\zb))\right)
\nonumber\\
&+\frac{1}{3} \left(
\text{Li}_3\left(\tfrac{\zb-z}{\zb}\right)
-\text{Li}_3\left(\tfrac{z-\zb}{z}\right)
+\text{Li}_3\left(\tfrac{z-\zb}{z-1}\right)
-\text{Li}_3\left(\tfrac{\zb-z}{\zb-1}\right)
+\text{Li}_3\left(\tfrac{z-\zb}{z(1-\zb)}\right)
-\text{Li}_3\left(\tfrac{\zb-z}{\zb(1-z)}\right)\right) \bigg]\,.
\nonumber
\end{align}
The first consistency check is that this correlator satisfies
\beq
F^{(2)} (z,\zb) = F^{(2)} (\zb,z)\,,
\label{eq:zzb_symmetry}
\eeq
a condition that is only satisfied because we input the spectrum of the CFT. If we let $\gamma_0^{(1)}$ unspecified and introduce a parameter $\gamma_\ell^{(1)} \to \alpha \gamma_\ell^{(1)}$, we find that \eqref{eq:zzb_symmetry} has two solutions corresponding to the Wilson-Fisher model and mean field theory. $\gamma_0^{(2)}$ is not constrained in this way because it multiplies a function of the form \eqref{eq:F1} which satisfies \eqref{eq:zzb_symmetry} by itself.
One can also expand in small $z$ and $\zb$ and find agreement with the known corrections to the OPE coefficients \cite{Gopakumar:2016cpb}\footnote{The expansion in small $z,\zb$ truncates the OPE sum over spin. The blocks for exchange of spinning operators in $d$ dimensions are given as a power series by the recursion relation of \cite{Dolan:2000ut}.}
\beq
a^{(2)}_{\ell} = \frac{(1+(-1)^\ell) \Gamma(\ell+1)^2}{\Gamma(2\ell+1)}
\frac{\ell (\ell+1) (H_{2\ell} - H_{\ell-1})-1}{9 \ell^2 (\ell+1)^2}\,,
\eeq
and \cite{Alday:2017zzv}
\beq
a^{MF}_{n,\ell} + \e^2 a^{(2)}_{n,\ell} = 
\begin{cases}
\frac{(1+(-1)^\ell)\Gamma(\ell+2)^2}{\Gamma(2\ell+3)}
\frac{\ell^2+3\ell+8}{24(\ell+1)(\ell+2)}
\left(
\frac{\e}{3}
\right)^2 + O(\e^3)\,, \quad &n=1\,,\\
O(\e^4)\,, \quad &n>1\,.
\end{cases}
\eeq

\subsection{Holographic supergravity}
\label{sec:sugra}

In order to give another example of applying the dispersion relation we will here apply it to compute the unprotected part of the reduced correlator $H(z,\zb)$ for the correlator $\langle \cO_2 \cO_2 \cO_2 \cO_2 \rangle$ in $\mathcal{N} = 4$ SYM theory at strong coupling, which holographically corresponds to the four graviton amplitude in AdS supergravity. While there is a vast literature on the topic, we recommend the beginning of \cite{Caron-Huot:2018kta} for a review of the setup and the definition of the reduced correlator.
From the usual bootstrap equation of the full correlator, the reduced correlator inherits the crossing equation
\bea
H(z,\zb) -H^{\text{free}}(z,\zb)={}& H(1-z,1-\zb) -H^{\text{free}}(1-z,1-\zb)\\
={}& (z \zb)^{-4} \left(H(1/z,1/\zb)-H^{\text{free}}(1/z,1/\zb) \right)\,,
\eea{eq:susy_bootstrap}
where
\beq
H^{\text{free}}(z,\zb) = \frac{1}{u^2} \left( 1+\frac{1}{v^2}
+\frac{1}{c} \frac{1}{v} \right)\,,
\label{eq:Hfree}
\eeq
is the reduced correlator of the free theory.
Moreover, it follows from the superconformal block expansion that $H(z,\zb)$ has the following expansion into the usual conformal blocks
\beq
 H(z,\zb) = (z \zb)^{-4} \sum\limits_{\De, \ell}  
a_{\De,\ell} \, g^{(4)}_{4+\De,\ell}(z,\bar{z})\,,
\label{eq:H_expansion}
\eeq
where the spectrum is given in an expansion around large central charge $c$ 
\beq
\De_{n,\ell} = 2n + \ell + \frac{1}{c} \gamma_{n,\ell}+ O(c^{-2})\,.
\eeq
Our final ingredient is that the operators with $n=0,1$ belong to short multiplets which are protected by supersymmetry. That means that 
their contribution does not receive perturbative corrections and can be computed once and for all 
\beq
H_{\text{short}} (z,\zb) = (z \zb)^{-4} \sum\limits_{n=0}^1 \sum\limits_{\substack{\ell=0\\\text{even}}}^\infty 
a_{n,\ell}  \, g^{(4)}_{4+2n+\ell,\ell}(z,\bar{z}) = (z \zb)^{-2} \mathcal{G}^{\text{short}} (z,\zb)  \,,
\eeq
with $\mathcal{G}^{\text{short}}$ as given in \cite{Beem:2016wfs}.
The goal of the following section is to derive the leading and subleading reduced correlator
\beq
H(z,\zb) = H^{(0)}(z,\zb) + \frac{1}{c}H^{(1)}(z,\zb)
= H_{\text{short}} (z,\zb) 
+ (z \zb)^{-4} \sum\limits_{n=2}^\infty \sum\limits_{\substack{\ell=0\\\text{even}}}^\infty
a_{n,\ell} \, g^{(4)}_{\De_{n,\ell},\ell}(z,\bar{z})
\,,
\label{eq:H_vs_Hshort}
\eeq
using as input only the anomalous dimensions and $H_{\text{short}}$, which contributes delta functions to the discontinuity due to its poles at $z=0$.

We will have to take into account that the 
presence of $H^{\text{free}}$ in the crossing equation \eqref{eq:susy_bootstrap} leads to an extra contribution in the dispersion relation. Repeating the argument of section \ref{sec:dispersion_relation}, the discontinuity at $z>1$ is now given by
\beq
\underset{z\, >\, 1}{\Disc}\, H(z,\zb) 
=- \underset{z\, <\, 0}{\Disc}\, \left( H(z,\zb)- H^{\text{free}}(z,\zb) + H^{\text{free}}(1-z,1-\zb) \right) \Big|_{\substack{z\to 1-z\\\zb \to 1-\zb}} \,,
\eeq
and the additional terms integrate to
\bea
h(z,\zb)&=
\frac{1}{2 \pi i} \int\limits_{-\infty}^0 \frac{dz'}{z' - (1-z)} \underset{z'}{\Disc}\, \left( - H^{\text{free}}(z',1-\zb) + H^{\text{free}}(1-z',\zb) \right) \\
&=\frac{-1}{((1-z)(1-\zb))^2} + \frac{1}{c}\frac{1-z-3 \zb +2z \zb}{((1-z)(1-\zb)\zb)^2}\,.
\eea{eq:h}
One checks from the $u$-channel OPE that the contribution from the arc at infinity vanishes for all exchanged operators and
is left with the following dispersion relation for the case at hand
\beq
H(z,\zb) = h(z,\zb) + \left(\frac{1}{2 \pi i} \int\limits_{-\infty}^0 dz' \, \frac{1}{z' - z} \underset{z'}{\Disc}\, H(z',\zb) + (z,\zb) \to (1-z,1-\zb)\right) \,.
\label{eq:bootstrap_dispersion_relation_H}
\eeq

\subsubsection{Leading order}

At order $c^0$ the discontinuity vanishes for $z<0$ due to \eqref{eq:disc_sin} and hence is located only at $z=0$ in the form of delta functions. From the OPE \eqref{eq:H_vs_Hshort} we see that poles at $z=0$ can only occur in $H_{\text{short}}$, which has the small $z$ expansion
\beq
H^{(0)}_{\text{short}} (z,\zb) =  \frac{2-2\zb+\zb^2}{(1-\zb)^2\zb^2 z^2} + \frac{2}{(1-\zb)^2\zb^2 z} + O(z^0)\,,
\eeq
and hence
\beq
\underset{z\, \leq\, 0}{\Disc}\, H^{(0)} (z,\zb) =
\underset{z\, \leq\, 0}{\Disc}\, H^{(0)}_{\text{short}} (z,\zb) =
2 \pi i \left(\frac{2-2\zb+\zb^2}{(1-\zb)^2\zb^2} \de'(z) - \frac{2}{(1-\zb)^2\zb^2} \de(z)\right)\,.
\label{eq:Disc_H0}
\eeq
Inserting this into the dispersion relation \eqref{eq:bootstrap_dispersion_relation_H} we obtain as expected
\beq
H^{(0)} (z,\zb) = \frac{1}{u^2} \left( 1 + \frac{1}{v^2} \right)\,.
\eeq

\subsubsection{Projecting out OPE coefficients}
\label{sec:projecting}

We use this opportunity to show a neat trick to quickly expand correlators for integer spectra into conformal blocks. Note that for some integer parameter $n$ the discontinuity at $z>1$ of the function $k_n(z)$ which appears in the definition of conformal blocks in $d=1,2,4,6,\ldots$ dimensions is given 
by Legendre polynomials $P_n(x)$
\beq
\underset{z\, \geq\, 1}{\Disc}\, k_{n} (z) = 2 \pi i \, r_n P_{n-1} \left( \frac{2-z}{z} \right)\,, \qquad n \in \mathbb{Z}^+\,, \qquad r_n = \frac{\Gamma(2n)}{\Gamma(n)^2}\,.
\label{eq:disc_k}
\eeq
Applying this twice to the conformal blocks appearing in
\beq
(z \zb)^{4} H^{(0)}(z,\zb) =  \sum\limits_{n=0}^\infty \sum\limits_{\substack{\ell=0\\\text{even}}}^\infty 
a^{(0)}_{n,\ell}\, g^{(4)}_{4+ 2n+\ell,\ell}(z,\bar{z})\,,
\label{eq:Hfree_expansion}
\eeq
one can also turn these blocks into orthogonal polynomials
\beq
\underset{\zb\, \geq\, 1}{\Disc}\, \underset{z\, \geq\, 1}{\Disc}\, 
g^{d=4}_{4+2n+\ell,\ell}(z,\bar{z}) =
(2 \pi i)^2 r_{n+1} r_{n+\ell+2} \frac{ z \bar{z}}{z-\bar{z}}
P_{n,\ell}^{-}\left(\frac{2-z}{z}, \frac{2-\zb}{\zb} \right)\,,
\label{eq:DiscG_4d}
\eeq
where we defined $P_{n,\ell}^{-}$ as the antisymmetric two variable Legendre polynomial
\beq
P_{n,\ell}^{-} (x, \xb) = P_{n+\ell+1} ( x ) P_{n} ( \xb ) - P_{n} ( x ) P_{n+\ell+1} ( \xb )\,,
\qquad n, \ell = 0,1,2,\ldots \,.
\label{eq:P_def}
\eeq
Using the orthogonality relation for ordinary Legendre polynomials, one finds that also the new polynomials obey an orthogonality relation
\beq
\int_{-1}^1 dx \int_{-1}^1 d \xb \, P_{n,\ell}^{-} (x, \xb) P_{n',\ell'}^{-} (x, \xb)
= \frac{8 \, \delta_{n n'} \delta_{\ell \ell'}}{(2n+1)(2(n+\ell+1)+1)} \,.
\label{eq:Legendre2_orthogonality}
\eeq
They form a basis for the antisymmetric two variable polynomials and have the properties
\beq
P_{n,\ell}^{-} (\xb, x) = - P_{n,\ell}^{-} (x, \xb)\,, \qquad
P_{n,\ell}^{-} (-\xb, -x) = (-1)^\ell P_{n,\ell}^{-} (x, \xb)\,.
\label{eq:Legendre2_symmetries}
\eeq
Given the discontinuity
\beq
\underset{\zb}{\Disc}\, \underset{z}{\Disc}\, (z \zb)^4 H^{(0)} (z,\zb) 
= (2 \pi i)^2
\left( 
\partial_{\frac{1}{z}} \delta(1/z)\partial_{\frac{1}{\zb}} \delta(1/\zb)
+
(z \zb)^{2} \partial_z \delta(1-z)\partial_{\zb} \delta(1-\zb)
\right)\,,
\eeq
we can directly compute all the OPE coefficients
\begin{align}
a^{(0)}_{n,\ell} ={}& \frac{(2n+1)(2(n+\ell+1)+1)}{2(2 \pi i)^2 r_{n+1} r_{n+\ell+2}} \int\limits_{1}^\infty \frac{dz}{z^2} \int\limits_{1}^\infty \frac{d\zb}{\zb^2} \, \frac{z-\zb}{z \zb}  P_{n,\ell}^{-} \left( \tfrac{2-z}{z}, \tfrac{2-\zb}{\zb} \right)
\underset{\zb}{\Disc}\, \underset{z}{\Disc}\, (z \zb)^4 H^{(0)}(z,\bar{z}) \nonumber\\
={}&  \frac{2(1+(-1)^\ell)(2n+1)(2(n+\ell+1)+1)}{r_{n+1} r_{n+\ell+2}}  
\partial_x P_{n,\ell}^{-} \left( x,\xb  \right)\Big|_{x=\xb=1}
\label{eq:anl_from_disc_F_4D}\\
={}& \frac{(1+(-1)^\ell)(2n+1)(2(n+\ell+1)+1)(\ell+1)(\ell+2n+2)}{r_{n+1} r_{n+\ell+2}}\,.
\nonumber
\end{align}
Unfortunately this trick is less useful at higher orders in the expansion, because then one needs to compute and subtract the term in \eqref{eq:HG_def_4D} which contains the anomalous dimensions before one can use the projection formula to compute $a^{(1)}_{n,\ell}$.

\subsubsection{Next-to-leading order}
Let us now consider the correlator at order $c^{-1}$. 
The anomalous dimensions for this correlator are known to be \cite{Alday:2014tsa}
\bea
\gamma_{n,\ell} = - \frac{(n-1)n(n+1)(n+2)}{(\ell+1)(\ell+2n+2)}\,.
\eea{eq:Dnl_N4}
The discontinuity for negative $z$ is given by \eqref{eq:DF_from_gamma}
\beq
\underset{z\, <\, 0}{\Disc}\, H^{(1)} (z,\zb) = \pi i (z \zb)^{-4} \sum\limits_{n=2}^\infty \sum\limits_{\substack{\ell=0\\\text{even}}}^\infty 
a_{n,\ell}^{(0)} \gamma_{n,\ell}  \, g^{(4)}_{4+2n+\ell,\ell}(z,\bar{z})\,.
\label{eq:DF_from_gamma_susy}
\eeq
While it is hard to do the sums in \eqref{eq:DF_from_gamma_susy}, one can check that with $a_{n,\ell}^{(0)}$ as in \eqref{eq:anl_from_disc_F_4D} the power series in $z, \zb$ agrees with\footnote{Note that ${\bar D}_{2422} (z,\zb) = \partial_u \partial_v (1+u\partial_u+v\partial_v) \frac{  \log \left( z \zb \right) \log \left( \frac{1-z}{1-\zb}\right) + 2 \text{Li}_2 (z) - 2 \text{Li}_2 (\zb) }{z-\zb}$.}
\begin{align}
&\underset{z\, <\, 0}{\Disc}\, H^{(1)} (z,\zb) = - \underset{z\, <\, 0}{\Disc}\, {\bar D}_{2422} (z,\zb)
\nonumber\\
{}&=\frac{8 \pi i}{(z-\zb)^7}  \bigg(3 \left(
2 (z^3 \zb + \zb^3 z + z^2+ \zb^2 )
-z^3-\zb^3 
+6 (z^2 \zb^2 + z \zb)
-9 (z^2 \zb + \zb^2 z )
\right) \text{log} \left(\frac{1-z}{1-\zb}\right)
\nonumber\\
&+\frac{1}{2 (\zb-1) (z-1)} 
\Big( 
(1-\zb) z^5 - (1-z) \zb^5
-28 (z^4 \zb^2 - \zb^4 z^2)
+39 (z^4 \zb - \zb^4 z)
-12 (z^4-\zb^4)
\nonumber\\
&+58 (z^3 \zb^2 - \zb^3 z^2)
-66 (z^3 \zb- \zb^3 z)
+12 (z^3 -\zb^3)
+24 (z^2 \zb- z \zb^2)
\Big) \bigg)\,.
\label{eq:Disc_Dbar2422}
\end{align}
However, in this case this is not the full discontinuity. 
Similar to the leading order, there are poles at $z=0$ from the protected operators
\beq
H^{(1)}_{\text{short}} (z,\zb) =  \frac{1}{(1-\zb)\zb^2 z^2} - \frac{2 (\zb + \log (1-\zb))}{\zb^4 z} + O(z^0)\,,
\eeq
causing the discontinuity
\beq
\underset{z\, \leq\, 0}{\Disc}\, H^{(1)}_{\text{short}} (z,\zb) =
2 \pi i \left(\frac{1}{(1-\zb)\zb^2 } \de'(z) + \frac{2 (\zb + \log (1-\zb))}{\zb^4} \de(z)\right)\,.
\label{eq:Disc_Hshort}
\eeq
Inserting the two contributions \eqref{eq:Disc_Dbar2422} and \eqref{eq:Disc_Hshort} to the discontinuity into the dispersion relation we compute the correct correlator \cite{Arutyunov:2000py,Dolan:2006ec}
\beq
H^{(1)}(z,\zb) = \frac{1}{u^2 v} - {\bar D}_{2422} (z,\zb)\,,
\eeq
from which the OPE coefficients $a^{(1)}_{n,\ell}$ can be extracted.

\section{Summary and Outlook}
\label{sec:5}

In this paper we have shown how to reconstruct four point correlators in generic number of dimensions from their discontinuity, which in perturbative settings is essentially determined by the spectrum of the theory. One of the main advantages of this method is that it allows us to compute the correlator without ambiguities, when the spectrum is specified. 

The method is general and can be applied to any CFT, in particular it would be very interesting to apply it to non-perturbative CFTs. Another idea is to use it for CFTs with boundaries or more general defects. In spirit it is very similar to \cite{Bissi:2018mcq}, which could be reformulated in terms of a dispersion relation as well. 

One could also explore the connection of this framework to the analytic functional approach \cite{Mazac:2016qev,Mazac:2018mdx,Mazac:2018ycv}. In the latter it is possible to extract the CFT data by constructing certain analytic functionals, named extremal functionals. 
Our section \ref{sec:projecting} could be reformulated in terms of functionals that extract the OPE coefficients from the discontinuity of a correlator in four dimensions, however these functionals have single poles, compared to the double poles of extremal functionals which can extract also the anomalous dimensions. Maybe our discussion can be useful for the construction of the full non-perturbative extremal functionals in 4D.

We conclude by mentioning another promising and interesting direction which is related to the Polyakov-Mellin bootstrap approach. It has been shown that there are ambiguities in this method which are due to the presence of contact terms which are not fixed by the Mellin bootstrap ansatz. Some progress in this direction has been made in \cite{Gopakumar:2018xqi}. We believe that our formulation could shed light on how to deal with and eventually remove such ambiguities. We hope to report on this problem in the near future.

\section*{Acknowledgments}

We thank Fernando Alday, Alessandro Georgoudis and Petr Kravchuk for helpful comments.
This research received funding from the Knut and Alice Wallenberg Foundation grant KAW 2016.0129, from VR grant 2018-04438 and was supported in part by Perimeter Institute for Theoretical Physics. Research at Perimeter Institute is supported by the
Government of Canada through the Department of Innovation, Science and Economic Development and by the Province of Ontario through the
Ministry of Research and Innovation.

\bibliographystyle{JHEP}
\bibliography{dispersion_relation}
\end{document}